\documentstyle[12pt]{article}

\renewcommand{\pd}{\partial}
\renewcommand{\a}{\alpha}
\renewcommand{\b}{\beta}
\newcommand{\ai}{\alpha_{I}}
\newcommand{\as}{\alpha_{\sigma}}
\newcommand{\aep}{\alpha_{\varepsilon}}
\newcommand{\s}{\sigma}
\newcommand{\ep}{\varepsilon}
\newcommand{\D}{\Delta}
\newcommand{\e}{{\rm e}}
\newcommand{\z}{{\bar z}}
\newcommand{\w}{{\bar w}}
\newcommand{\half}{\frac{1}{2}}
\newcommand{\fr}{\frac}
\newcommand{\sq}{{\sqrt 6}}
\newcommand{\Oi}{O_I}
\newcommand{\Os}{O_{\sigma}}
\newcommand{\Oep}{O_{\varepsilon}}
\newcommand{\zp}{z^{\prime}}
\renewcommand{\wp}{w^{\prime}}
\newcommand{\pp}{\prime}
\newcommand{\Bs}{B_{\sigma}}
\newcommand{\Rs}{R_{\sigma}}
\newcommand{\Bep}{B_{\varepsilon}}
\newcommand{\Rep}{R_{\varepsilon}}
\newcommand{\bb}{\begin{equation}}
\newcommand{\ee}{\end{equation}}
\newcommand{\bba}{\begin{eqnarray}}
\newcommand{\eea}{\end{eqnarray}}

\begin{document}

\begin{titlepage}

\begin{tabbing}
   qqqqqqqqqqqqqqqqqqqqqqqqqqqqqqqqqqqqqqqqqqqqqq
   \= qqqqqqqqqqqqq  \kill
         \> {\sc KEK-TH-417} \\
         \> {\sc November 1994}
\end{tabbing}
\vspace{5mm}

\begin{center}
{\Large {\bf Recursion Relations in Liouville Gravity
 coupled to Ising Model satisfying \break  Fusion Rules}}
\end{center}

\vspace{1.5cm}

\centering{\sc Ken-ji HAMADA}\footnote{E-mail address : hamada@theory.kek.jp}

\vspace{1cm}

\begin{center}
{\it National Laboratory for High Energy Physics (KEK),} \\
{\it Tsukuba, Ibaraki 305, Japan}
\end{center}

\vspace{1cm}

\begin{abstract}
The recursion relations of 2D quantum gravity coupled to the Ising
model discussed by the author previously are reexamined.
We study the case in which the matter sector satisfies the fusion rules
and only the primary operators inside the Kac table contribute. The
theory involves unregularized divergences in some of correlators.
We obtain the recursion relations which form a closed set among
well-defined correlators on sphere, but they do not have
a beautiful structure that the bosonized theory has and also give an
inconsistent result when they include an ill-defined correlator with  the
divergence. We solve them and compute the several normalization
independent ratios of the well-defined correlators,
which agree with the matrix model results.
\end{abstract}
\end{titlepage}

\section{Introduction}
\indent

   In conformal gauge, 2D quantum gravity is described
as the Liouville theory coupled to matter field [1--11].
The Liouville field gives the gravitational dressing to ensure the
general covariance. As a matter sector we take a conformal field
theory (CFT). In general CFT has the finite number
of primary fields and they satisfy the fusion rules~\cite{bpz}.
However, in the matrix model approach
the violation of the fusion rules is observed~\cite{cgm,dfk}.
This problem is overcome by using the bosonized fields for the
matter sector~\cite{gl,hb}. In this case, as discussed by Kitazawa in
ref.~\cite{gl}, the zero of the correlation function for the matter sector,
which represents the fusion rule, is canceled with the
divergence of the Liouville sector so that the combined correlaton function
becomes finite. And also the primary operators
outside the Kac table no longer decouple, which are identified with the
gravitational descendants. We can also derive the non-linear structures called
W algebra constraints~\cite{fkn} as the Ward identities~\cite{hb} of
the $W_{\infty}$ symmetry~\cite{w}. The formalism can be easily extended to 2D
quantum supergravity~\cite{hi}.  Anyway the bosonized theory has
the same beautiful structures as what the matrix model has.

  At the first stage of 2D quantum gravity [1--5], however,
we mainly considered the usual CFT satisfying the fusion rules and reached
the good agreement with the matrix model results.
Here we reconsider the situation that the
fusion rules are maintained and observe how well such a agreement is.

  As a concrete example to study such an issue we take up the Liouville
gravity coupled to the Ising model. The Ward identities of this system have
been discussed by the author previously~\cite{ha}.
Here we reexamine and make clear the previous arguments and complete the
calculations. Some of correlators in this theory become ill-defined due to
appearance of unregularized divergences.  The Ward identity is then
well-defined if all correlators involved in it are well-defined,
but gives an inconsistent result when an ill-defined correlator appears
in the expression.

   The Ward identities of the system satisfying the fusion rules are
rather tricky. In the bosonized theory there exists the BRST invariant
$W_{\infty}$ current~\cite{w,bmp},
but in the non-bosonized theory such a current dose not exist.  So we here
use a non-primary current like $\pd \phi$, where $\phi$ is
the Liouville field, which leads to a complicated recursion relation.
The derived recursion relation has a similar structure to the W algebra
constraint~\cite{fkn}. We can regard it as a deformed W algebra constraint
due to keeping the fusion rules and
setting the gravitational descendants located outside Kac table decoupled.

   A closed set of recursion relations on sphere is obtained. They have
the three independent parameters related to the normalizations of three
scaling operators and explicitely depend on how to assign curvature
singularities
which are here put on the scaling operators. But the solutions of them are
independent of how to choose the assignment.
We obtain some solutions and calculate
the normalization independent ratios. The results agree with those calculated
in the other methods~\cite{cgm,dfk,gl}.

\section{Quantum Liouville Theory coupled to Ising Model}
\indent

  Two dimensonal quantum gravity is defined through the functional integrations
over metric tensor of 2D surface $g_{\a\b}$ and matter fields. After
fixing the reparametrization invariance in conformal gauge
$g =\e^{\ai \phi}{\hat g}$, where $\phi$ is the Liouville field and
${\hat g}$ is the background metric, the theory is expressed as the combined
Liouville, matter and ghosts system defined by the action
$S=S^L + S^M + S^G $. Here
$S^L$ is the Liouville one
\bb
   S^L = \fr{1}{8\pi}\int d^2 z \sqrt{{\hat g}}
         \bigl( {\hat g}^{\a\b} \pd_{\a}\phi  \pd_{\b}\phi
                  +Q{\hat R} \phi \bigr)
             + \mu \int d^2 z \e^{\ai\phi}
\ee
and $S^M$ and $S^G$ are the matter and the $bc$-ghosts ones.
In this paper we argue only the Ising model as a matter sector for
simplicity. In this case we can derive the concrete results.
The parameters $Q$ and $\ai$ are then
\bb
      Q=\fr{7}{\sq} ~, \qquad \ai =\fr{3}{\sq} ~.
\ee

  The Ising model has the three primary states with conformal weights
$\Delta =0$, $\fr{1}{16}$, $\fr{1}{2}$. The fusion rules ,or operator
product expansions (OPE) of the primary fields are
\bba
 && \ep(z,\z) \ep(0,0) = \fr{1}{|z|^2} ~,
             \nonumber \\
 && \ep(z,\z) \s (0,0) = \fr{C_{\ep\s\s}}{|z|} \s (0,0) ~, \\
 && \s (z,\z) \s (0,0) = \fr{1}{|z|^{1/4}}
                           + C_{\ep\s\s}|z|^{3/4} \ep (0,0) ~,
                \nonumber
\eea
where $\ep(z,\z)$ is the energy density operator with weight
$(\Delta, {\bar \Delta})=(\half,\half)$ and $\s (z,\z)$ is the spin field
with weight $(\fr{1}{16}, \fr{1}{16})$. The OPE coefficient is given by
$C_{\ep\s\s}=\half$~\cite{dfsz}. Throughout paper we keep the above OPE's.

   The physical scaling operators are now given by dressing the matter
primary fields $I$, $\s$ and $\ep$, where $I$ denotes the identity, which
are
\bba
  &&  \Oi = \int d^2 z \e^{\ai \phi(z,\z)} ~, \qquad
     \Os = \int d^2 z \e^{\as \phi(z,\z)} \s(z,\z) ~,
         \nonumber    \\
  &&  \Oep = \int d^2 z \e^{\aep \phi(z,\z)} \ep (z,\z) ~,
\eea
where $\as = \fr{5}{2\sq}$ and $\aep = \fr{1}{\sq}$. $\ai$ is given in
eq.(2.2). For the sake of subsequent discussion we introduce the notations
\bb
     O_a (z,\z)= {\bar c}(\z)c(z)V_a (z,\z) ~, \qquad
      O_a = \int d^2 z V_a (z,\z) ~,
\ee
where $a=I,\s,\ep$.

   Let us define the correlation functions of the Liouville gravity.
According to the argument of Goulian-Li in ref.~\cite{gl},
we first integrate over the zero
mode of the Liouville field. Then the correlation function is expressed by
the free field one
\bb
   \ll {\cal O} \gg_g = \kappa^{-\chi} \mu^s
        \fr{\Gamma(-s)}{\ai} < {\cal O}~ \bigl( \Oi \bigr)^s >_g ~,
\ee
where $g$ is genus of 2D surface and $\chi= 2-2g$. The $\Gamma$-function
comes from the zero mode integral of $\phi$. For
\bb
     {\cal O}=\prod^{n_1} \Oi \prod^{n_2} \Os \prod^{n_3} \Oep
\ee
$s$ is given by
\bba
   &&  s = \fr{1}{\ai} \biggl[
           \fr{Q}{2}\chi -n_1 \ai -n_2 \as -n_3 \aep
                     \bigg]   \nonumber  \\
   && \quad = \fr{7}{6}\chi -n_1 -\fr{5}{6} n_2 -\fr{1}{3} n_3  ~.
\eea
The expression is convergent for $s < 0$, while for $s \geq 0$ the integral
diveges. But the $\Gamma$-function can analytically continue to the region
$s>0 ~(s \neq \mbox{integer})$ so that  the expression is considered as a
regularized form for $s>0$. This ensures that the correlator satisfies the
relation $-\fr{\pd}{\pd \mu} \ll {\cal O} \gg_g = \ll \Oi {\cal O} \gg_g$
even though $s >0 ~(s \neq {\bf Z}_+ )$.
For the cases that $s$ is zero or positive
integer the correlation function becomes ill-defined.
To overcome this problem we have to go to the bosonized theory perturbed by
the cosmological constant operator and one of the screening charges for
matter sector~\cite{hb}.

\section{Equation of Motion}
\setcounter{equation}{0}
\indent

  We first consider the current $\pd \phi(z)$.  The Ward identity is then
\bb
   \int d^2 z {\bar \pd} \ll \pd \phi (z) ~{\cal O} \gg_g =0 ~.
\ee
The OPE between $\pd \phi$ and the scaling operator $O_a$
$( a=I,~\s,~\ep )$ is naively caluculated as
$\pd \phi(z) O_a(w) =-\fr{\a_a}{z-w} O_a (w)$.

 If the operator $\pd \phi$ were primary, there would be no problem.
However it is not primary, which is subject to
$L_1 \cdot \pd \phi = Q$ and $L_n \cdot \pd \phi = 0 ~(n \geq 2)$, such
that it is transformed as
\bb
   \pd \phi (z) = \biggl( \fr{\pd \zp}{\pd z} \biggr)
                    \biggl[  \pd^{\pp} \phi^{\pp}(\zp)
                  -\fr{Q}{2}  \fr{\pd^2 z}{\pd z^{\pp 2}}
                           \fr{\pd \zp}{\pd z} \biggr] ~.
\ee
Then we have to pay attention to the curvature singularity. Here we
consider the case that the background metric which is almost flat except
for the $\delta$-function singularities at the positions of the scaling
operators as~\cite{p}
\bb
       \fr{1}{4\pi} \sqrt{\hat g}{\hat R}
          = \sum_a \nu_a \delta^2 (z-z_a)
\ee
such that $\sum_a \nu_a = \chi$, where $\chi=2-2g$ is the Euler number.
We then do not assign the curvature on $\pd \phi$ and the potential
term.

   When we calculate the OPE we need to smooth out the curvature
singularity in the neighborhood of the position of the scaling
operator, which can be carried out by using the transformation
\bb
    z - z_a = (\zp - \zp_a )^{1-\nu_a} ~, \quad \mbox{or}
     \quad dz d\z = \fr{d\zp d{\bar \zp}}{|\zp - \zp_a |^{2\nu_a}}~.
\ee
After going to the smooth $\zp$-frame, the OPE between
$\pd \phi (z)$ and $O_a (z_a, \z_a)$ is calculated as follows:
\bba
    dz \pd \phi(z) O_a (z_a, \z_a) &=&
        d\zp \biggl( \pd^{\pp} \phi^{\pp} (\zp)
          + \fr{Q}{2}\nu_a \fr{1}{\zp-\zp_a} \biggr)
             O^{\pp}_a (\zp_a, {\bar \zp}_a)
              \nonumber \\
      &=& \biggl( -\a_a + \fr{Q}{2}\nu_a \biggr) d\zp
              \fr{1}{\zp-\zp_a} O^{\pp}_a (\zp_a, {\bar \zp}_a) ~.
\eea

  The derivative ${\bar \pd}$ in eq.(3.1) picks up the OPE singularity.
For the case (2.7) we get the following expression:
\bb
   \biggl( \fr{Q}{2} \chi  -n_1 \ai -n_2 \as -n_3 \aep
            -s \ai \biggr) \ll {\cal O} \gg_g  = 0 ~.
\ee
Here we use the relation $\sum_a \nu_a = \chi $.
This equation is trivial, which nothing but say that $s$ is given by
eq.(2.8). This equation stands for that the Liouville field satisfy the
equation of motion
\bb
    {\bar \pd} \pd \phi(z,\z) = \pi \ai \mu \e^{\ai \phi (z,\z)}
\ee
in the interaction picture $\ll \cdots \gg$.

   Here the curvature treatment does well for arbitrary genus, but
in general it breaks down for higher genus because the coordinate
transformation (3.4) is not globally well-defined for higher genus.
In the sequence we consider only the sphere topology.

\section{Recursion Relations}
\setcounter{equation}{0}
\indent

  In this section we consider more complicated Ward identities which give
informations for correlation functions.  Let us introduce the operator
of the form $W (z,\z) = R(z) {\bar B}(\z)$, where $R(z)$ has the conformal
weight 1 and $B(z)$ has the weight 0 and they have the same Liouville
charge. As non-trivial operators of $B(z)$, we here consider the following
BRST invariant ones:
\bb
   \Bs (z) = \biggl( bc\s - \fr{4}{2\sq}\pd\phi\s
                              -\fr{4}{3}\pd\s   \biggr)
                   \e^{-\fr{3}{2\sq} \phi (z)}
\ee
and
\bb
   \Bep (z) = \biggl( bc\ep - \fr{3}{2\sq}\pd\phi\ep
                              -\fr{3}{4}\pd\ep   \biggr)
                   \e^{-\fr{4}{2\sq} \phi (z)} ~,
\ee
which correspond to the ring elements in the bosonized theory [8--10].
The divergence of $W(z,\z)$ is
\bb
      {\bar \pd} W(z,\z) = R(z) \{ {\bar Q}_{BRST},
                  [{\bar b}_{-1}, {\bar B}(\z)] \} ~,
\ee
where we use the fact that ${\bar B}(\z)$ is the BRST invariant operator.
The anti-holomorphic part is BRST-trivial so that the operator
$W(z,\z)$ plays a role of current.

  The operator $R(z)$ has the same Liouville charge as $B(z)$ and the
conformal weight 1. Therefore the partners of $\Bs(z)$ and $\Bep(z)$ should
include the two derivatives. Fitting in our previous work~\cite{ha},
we here take up the following forms:\footnote{
We may use the other type of the operator because we here do not impose
the primary conditions.} 
\bb
    \Rs (z) = \biggl( \fr{7}{3} \pd^2 \phi
                       - \fr{4}{\sq} (\pd \phi )^2 \biggr)
                    \e^{-\fr{3}{2\sq}\phi}\s (z)
\ee
and
\bb
    \Rep (z) = \biggl( \fr{7}{4} \pd^2 \phi
                       - \fr{3}{\sq} (\pd \phi )^2 \biggr)
                    \e^{-\fr{4}{2\sq}\phi}\ep (z) ~.
\ee
These are not the primary operators. In the bosonized theory
there exist the primary states, callled the discrete states,
which form the $W_{\infty}$ algebra. In this paper, however, we do not
use the bosonization for the matter sector. Then such a primary state
does not exist so that in this theory the Ward identities do not have
the beautiful structures like the W algebra constraints.
As discussed below the Ward identities have more complicated forms and
we cannot argue them beyond the topology of sphere.

  The complexity comes from the non-tensor transformation property of
$R(z)$. The operator $\Rs(z)$ satisfies the equations
\bba
   &&  L_1 \cdot \Rs(z) = S_{\s}(z) ~,
          \quad L_2 \cdot \Rs(z)=\fr{9}{4}T_{\s}(z) ~,
                      \\
   &&           L_1 \cdot S_{\s}(z)=-\fr{17}{8} T_{\s}(z) ~,
\eea
$L_n \cdot \Rs(z) =0 ~(n \geq 3)$ and $L_n \cdot S_{\s}(z)=0~(n \geq 2)$,
where $L_n = L^L_n + L^M_n + L^G_n$ and
\bb
      L^{L,M,G}_n \cdot A(z)
        = \oint_{C_z} \fr{dy}{2\pi i}(y-z)^{n+1}
             T^{L,M,G}(y) A(z)
\ee
for some operator $A(z)$. Here $T^{L,M,G}(z)$ are the energy-momentum
tensors for the Liouville, the matter and the ghost sectors respectively.
These imply that $\Rs(z)$ is transformed as
\bba
     && \Rs(z) = \biggl( \fr{\pd \zp}{\pd z} \biggr)
            \biggl[ \Rs^{\pp}(\zp)
              -\half \fr{\pd^2 z}{\pd z^{\pp 2}}
                   \fr{\pd \zp}{\pd z}S_{\s}^{\pp}(\zp)
              -\fr{3}{8} \fr{\pd^3 z}{\pd z^{\pp 3}}
                    \fr{\pd \zp}{\pd z} T_{\s}^{\pp}(\zp)
              \nonumber     \\
    && \qquad\qquad\qquad\qquad\qquad
             +\fr{19}{64} \biggl(
                  \fr{\pd^2 z}{\pd z^{\pp 2}} \biggr)^2
                  \biggl( \fr{\pd \zp}{\pd z} \biggr)^2
                    T_{\s}^{\pp}(\zp)  \biggr] ~,
\eea
where $S_{\s}(z)$ and $T_{\s}(z)$ are the operators with weights $0$
and $-1$ respectively defined by
\bb
   S_{\s}(z)=L_{-1}^{L} \cdot T_{\s}(z) ~, \qquad
        T_{\s}(z) = \fr{80}{3\sq}\e^{-\fr{3}{2\sq}\phi}\s(z)
\ee
which is transformed as
\bba
     && S_{\s}(z) = S_{\s}^{\pp}(\zp)
            + \fr{17}{16}\fr{\pd^2 z}{\pd z^{\pp 2}}
                   \fr{\pd \zp}{\pd z} T_{\s}^{\pp}(\zp) ~,
                        \\
     && T_{\s}(z) = \biggl( \fr{\pd \zp}{\pd z} \biggr)^{-1}
                        T_{\s}^{\pp}(\zp) ~.
\eea

  The transformation law of the operator $R_{\ep}(z)$ is also given
as follows:
\bba
     && \Rep(z) = \biggl( \fr{\pd \zp}{\pd z} \biggr)
            \biggl[ \Rep^{\pp}(\zp)
              -\half \fr{\pd^2 z}{\pd z^{\pp 2}}
                   \fr{\pd \zp}{\pd z}S_{\ep}^{\pp}(\zp)
              -\fr{17}{36} \fr{\pd^3 z}{\pd z^{\pp 3}}
                    \fr{\pd \zp}{\pd z} T_{\ep}^{\pp}(\zp)
              \nonumber     \\
    && \qquad\qquad\qquad\qquad\qquad
             +\fr{1}{3} \biggl(
                  \fr{\pd^2 z}{\pd z^{\pp 2}} \biggr)^2
                  \biggl( \fr{\pd \zp}{\pd z} \biggr)^2
                    T_{\ep}^{\pp}(\zp)  \biggr] ~,
\eea
where
\bba
   &&  L_1 \cdot \Rep(z) = S_{\ep}(z) ~,
          \quad L_2 \cdot \Rep(z)=\fr{17}{6}T_{\ep}(z) ~,
                      \\
   &&           L_1 \cdot S_{\ep}(z)=-3 T_{\ep}(z)
\eea
and
\bb
   S_{\ep}(z)=L_{-1}^{L} \cdot T_{\ep}(z) ~, \qquad
        T_{\ep}(z) = \fr{33}{2\sq}\e^{-\fr{4}{2\sq}\phi}\ep(z) ~.
\ee

\subsection{Ward identities for $W_{\s}$}
\indent

   We first consider the Ward identities for the current
$W_{\s}(z,\z)$, which is
\bb
      \int d^2 z {\bar \pd} \ll W_{\s}(z,\z)~ {\cal O}\gg_0 =0
\ee
Let us calculate the OPE between the current and the scaling operators.
As discussed in Sect.3 the curvature singularities are assigned on the
scaling operators and we treat them with care.
By transforming into the non-singular frame we get the following OPE:
\bba
   &&  dz W_{\s}(z,\z) \Os (w,\w)
         \nonumber     \\
   && \qquad  = d \zp \biggl[
          R^{\pp}_{\s}(\zp)
           +\fr{\nu}{2}\fr{1}{\zp -\wp}S^{\pp}_{\s}(\zp)
           - \fr{5\nu^2 +24\nu}{64(\zp -\wp)^2} T^{\pp}_{\s}(\zp)
                 \biggr]
               \nonumber \\
   && \qquad\qquad\qquad\qquad\qquad \times
        {\bar B}^{\pp}_{\s}(\z^{\pp})\Os^{\pp}(\wp, \w^{\pp})
             \nonumber   \\
   && \qquad  = d \zp \fr{1}{\zp-\wp}
           \biggl( - \fr{10}{9\sq}\biggr)  C_{\ep\s\s}
            \biggl( 1-\nu-\fr{5}{4} \nu^2 \biggr)
              \Oep^{\pp}(\wp, \w^{\pp}) ~,
\eea
where $\nu$ is the curvature singularity
assigned on $\Os$. In order to smooth away it we use the coordinate
transformation (3.4) and the transformation law of $R_{\s}(z)$ (4.9).
The OPE's of the current with $\Oi$ and $\Oep$ do not give any
contribution.

  The derivative ${\bar \pd}$ in eq.(4.16) picks up the OPE singularity
calculated above and also produce the following correlator:
\bb
    \ll \int d^2 z {\bar \pd}W_{\s}(z,\z) ~{\cal O} \gg_0 ~,
\ee
where the anti-holomorphic part of ${\bar \pd}W_{\s}(z,\z)$ is
BRST-trivial (4.3). So this correlator usually would vanish.
However, as discussed in the previous papers~\cite{p,ha,hb}, the boundary
of moduli space is now dangerous and this correlator gives an
anomalous contribution.

  Using the factorization formula, we can calculate such a contribution.
We need to evaluate the following expression:
\bba
    && \kappa^2 \sum^{\infty}_{N=1}\sum_{\D}
           \int^{\infty}_{-\infty} \fr{dp}{2\pi}
             \ll {\cal O}_1
               \int_{|z| \leq 1} d^2 z {\bar \pd}W_{\s}(z,\z)
             \nonumber \\
    && \qquad\qquad\qquad
           \times D |-p, \D ; N \gg_0 \ll p, \D; N| {\cal O}_2 \gg_0
\eea
where $\D = 0$, $\fr{1}{16}$ and $\half$. $D$ is the propergator.
The intermediate state is
the normalizable eigenstate of the Hamiltonian $H=L_0 +{\bar L}_0$
with the eigenvalue $p^2 +\fr{Q^2}{4} +2(\D +N-1)$. For $N=0 $ it is
given by
\bb
    |p, \D> = \e^{(ip +\fr{Q}{2})\phi(0,0)}\Phi_{\D}(0,0)
                |0>_{L,M} \otimes {\bar c}_1 c_1 |0>_G ~,
\ee
where $\Phi_{\D} =I, \s, \ep$. $N \neq 0$ states are their descendants.
The normalization is
\bb
     \ll p^{\pp},\D^{\pp};N^{\pp} | p,\D;N \gg_0
        = \kappa^{-2} 2\pi \delta(p+p^{\pp}) \delta_{\D,\D^{\pp}}
                 \delta_{N,N^{\pp}} ~.
\ee
The zero mode integral of the Liouville field now produces the
$\delta$-function.

  The propagator is defined by
\bb
    D = \int_{|z| \leq 1} \fr{d^2 z}{|z|^2} z^{L_0}\z^{{\bar L}_0}
      = 2\pi \biggl( \fr{1}{H} - \lim_{\tau \rightarrow \infty}
                       \fr{1}{H} \e^{-\tau H} \biggr) ~,
\ee
where we introduce the regulator at $|z|=\e^{-\tau}$. The last term
stands for the boundary of moduli space. Since the BRST charge
commutes with the Hamiltonian there is no contribution from
$1/H$ term. The boundary term is now dangerous. So we evaluate
the following quantity carefully:
\bba
   && \lim_{\tau \rightarrow \infty} \kappa^2  \sum_{\D}
         \int^{\infty}_{-\infty} \fr{dp}{2\pi}
          \int_{\e^{-\tau} \leq |z| \leq 1} d^2 z
         \ll {\cal O}_1 ~
               R_{\s}(z)[{\bar b}_{-1}, {\bar B}_{\s}(\z)]
             \nonumber     \\
   && \qquad\quad \times
        {\bar Q}_{BRST}
              \biggl( -\fr{2\pi}{H}\e^{-\tau H} \biggr)
            |-p, \D; N \gg_0 \ll p,\D; N| {\cal O}_2 \gg_0 ~,
\eea
where we omit $N \neq 0$ modes because, as discussed below, these
modes vanish at $\tau \rightarrow \infty$. Noting
\bb
    {\bar Q}_{BRST} \biggl( -\fr{2\pi}{H} \e^{-\tau H}
              \biggr) |-p, \D >
      = -\pi {\bar \pd}{\bar c}(0)\e^{-\tau H}
                  |-p, \D >
\ee
and
\bb
        [ {\bar b}_{-1}, {\bar \Bs}(\z) ]
         = -{\bar b}(\z) \e^{-\fr{3}{2\sq}\phi(\z)}\s(\z) ~,
\ee
we obtain the expression
\bba
  && \lim_{\tau \rightarrow \infty} \kappa^2 \sum_{\D}
         \int^{\infty}_{-\infty} \fr{dp}{2\pi}
            \pi C_{\s \D \D^{\pp}}
              A_{\s}(p,\D) \e^{-\tau(p^2 +\fr{Q^2}{4}+2\D-2)}
               \nonumber  \\
  && \qquad\qquad \times
       \int_{\e^{-\tau} \leq |z| \leq 1} d^2 z
          |z|^{2\{-2 +\fr{3}{2\sq}(-ip +\fr{Q}{2})
                   -\fr{1}{16}-\D+\D^{\pp} \}}
              \nonumber   \\
  && \qquad\qquad  \times
       \ll {\cal O}_1 |-p+i\mbox{$\fr{3}{2\sq}$}, \D^{\pp} \gg_0
         \ll p, \D | {\cal O}_2 \gg_0 ~,
\eea
where $C_{\s \D \D^{\pp}}$ is the OPE coefficient for the matter sector
defined by eq.(2.3).
When the correlator is divided into two on sphere, the curvature
singularity is induced at the state $|-p,\D \gg$ to ensure that the Euler
number is $2$ for sphere. We here describe the induced curvature on the
state as $\nu_p$. Simultaneously the curvature $2-\nu_p$ is induced at
the state $\ll p, \D|$.
The coefficient $A_{\s}(p,\D)$, which comes from the Liouville sector,
 is given as follows:
\bba
   &&  A_{\s}(p,\D) =
          \fr{7}{3} \biggl(-ip+\fr{Q}{2} \biggr)
            -\fr{4}{\sq} \biggl( -ip+\fr{Q}{2} \biggr)^2
                  \nonumber \\
   && \qquad\qquad\qquad
           +\fr{10}{3} \nu_p \biggl(-ip+\fr{Q}{2} \biggr)
             -\fr{80}{3\sq}\fr{5\nu^2_p +24\nu_p}{64} ~.
\eea

   Changing the variable to $z = \e^{-\tau x +i\theta}$, where
$0 \leq x \leq 1$ and $0 \leq \theta \leq 2\pi$, the above expression
is rewritten as
\bba
   && \lim_{\tau \rightarrow \infty} \kappa^2 \sum_{\D}
         \int^{\infty}_{-\infty} \fr{dp}{2\pi}
           \int^1_0 2\pi\tau dx ~\pi C_{\s\D\D^{\pp}}~ A(p,\D)
             \nonumber  \\
   &&  \times
          \ll {\cal O}_1 |-p+i\mbox{$\fr{3}{2\sq}$}, \D^{\pp} \gg_0
         \ll p, \D | {\cal O}_2 \gg_0
             \exp \biggl[ -\tau \Bigl\{
                 p^2 + \mbox{$\fr{1}{24}$}+2\D
               \nonumber  \\
   && \qquad\qquad
               +2x\Bigl( \mbox{$\fr{3}{2\sq}$}
                        \Bigl(-ip+\mbox{$\fr{Q}{2}$} \Bigr)
                  -1 -\mbox{$\fr{1}{16}$}-\D +\D^{\pp} \Bigr)
                         \Bigr\} \biggr] ~.
\eea
Since the exponential term is highly peaked in the limit
$\tau \rightarrow \infty$, the saddle point estimation becomes exact.
The saddle point of the $p$ integral is $p = i\fr{3}{2\sq}x$, so that
(4.29) becomes
\bba
    && \lim_{\tau \rightarrow \infty} \kappa^2 \sum_{\D}
             \pi \tau  \sqrt{\fr{2\pi}{2\tau}}
                  \int^1_0 dx ~ C_{\s\D\D^{\pp}}
                   ~ A \Bigl( \mbox{$i\fr{3}{2\sq}x$}, \D \Bigr)
                   \nonumber  \\
    && \qquad \times
           \ll {\cal O}_1 |i\mbox{$\fr{3}{2\sq}$}(1-x), \D^{\pp} \gg_0
         \ll i\mbox{$\fr{3}{2\sq}$}x , \D | {\cal O}_2 \gg_0
                  \nonumber  \\
    && \qquad \times
          \exp \biggl[ -\tau \Big\{ \mbox{$\fr{3}{8}$}x^2
                -\Bigl( \mbox{$\fr{3}{8}$} +2\D -2\D^{\pp} \Bigr)x
                 + \mbox{$\fr{1}{24}$}+2\D \Bigr\} \biggr]  ~.
\eea
$x$ integral is also evaluated  at the saddle point
\bb
     x = \fr{1}{6} (3 + 16\D - 16\D^{\pp} ) ~.
\ee
{}From the fusion rules (2.3) the possible pairs of $(\D, \D^{\pp})$ are
$(0, \fr{1}{16})$, $(\half,\fr{1}{16})$, $(\fr{1}{16},0)$ and
$(\fr{1}{16}, \half)$. In order to get non-vanishing contributions it
is necessary that the saddle point is located within the interval
$0 \leq x \leq 1$. Thus only two cases $(0, \fr{1}{16})$ and
$(\fr{1}{16},0)$ survive. For $(0, \fr{1}{16})$ the saddle point is
$x= \fr{1}{3}$ and we obtain
\bb
      -\pi^2 \kappa^2 \fr{8}{9}
        \biggl( \fr{5}{4}\nu_p -1 \biggr)^2
          \ll {\cal O}_1 ~ \Os \gg_0
           \ll \Oi ~ {\cal O}_2  \gg_0 ~.
\ee
For $(\fr{1}{16},0)$ the saddle point is $x=\fr{2}{3}$ and we get
\bb
      -\pi^2 \kappa^2 2\biggl( \fr{5}{6}\nu_p -1 \biggr)^2
        \ll {\cal O}_1 ~ \Oi \gg_0
           \ll \Os ~ {\cal O}_2  \gg_0 ~.
\ee

   Replacing $\D$ and $\D^{\pp}$ with $\D+N $ and $\D^{\pp}+N$ in the
expression (4.29), we can see that the contributions from oscilation modes
vanish exponentially as $\e^{-2N\tau}$.
Combining the results (4.32) and (4.33) we obtain the following expression:
\bba
   && \ll \int d^2 z {\bar \pd} W_{\s}(z,\z)
             \prod_{a \in S} O_a  \gg_0 =
                  \nonumber  \\
   &&  -\fr{\pi^2 \kappa^2}{2} \sum_{S=X \cup Y} \biggl[
        \fr{8}{9} \biggl( \fr{5}{4}\nu_p -1 \biggr)^2
        \ll  \Os ~\prod_{b \in X}O_b  \gg_0
           \ll \Oi ~ \prod_{c \in Y} O_c \gg_0
                  \nonumber  \\
   && \qquad\qquad
          + 2 \biggl( \fr{5}{6}\nu_p -1 \biggr)^2
        \ll \Oi ~\prod_{b \in X}O_b  \gg_0
           \ll \Os  \prod_{c \in Y} O_c \gg_0
                      \biggr] ~,
\eea
where the factor $\half$ in r.h.s. corrects for double counting.
The induced curvature $\nu_p$ satisfies the relation
\bb
             \nu_p + \sum_{b \in X}\nu_b = 2 ~.
\ee
Note that $\nu_p$ also satisfies the relation
$2-\nu_p + \sum_{c \in Y}\nu_c = 2$ because of
$\sum_{a \in S}\nu_a =2$. The first and the second terms of the r.h.s.
of eq.(4.34) are symmetric under the interchange
$\nu_p \leftrightarrow 2-\nu_p$.

   Let us first consider the case of ${\cal O}=\prod^n \Os $.
To write down the expression we have to specify the assignment of the
curvature singularity. The expression depends on the assignment, but
the solution is independent of how to choose it. Here we properly assign
the curvatures $\nu_a ~(a=1 \cdots n)$ on $n$ $\Os$ operators, where
$\sum^n_{a=1}\nu_a =\chi=2$. Then we get
\bba
   && \pi \biggl( -\fr{10}{9\sq} \biggr) C_{\ep\s\s}
          \biggl( n-2-\fr{5}{4} \sum^n_{a=1} \nu^2_a \biggr)
        \ll \Oep \prod^{n-1} \Os \gg_0
               \nonumber  \\
   && -\fr{\pi}{\sq} \lambda_1  \sum^n_{k=0}
            \biggl[ 2 {n-2 \choose k} + \fr{8}{9}{n-2 \choose k-2}
                  -\fr{8}{3}{n-2 \choose k-1}
                           \\
   && \qquad
               + \fr{25}{18} {n-2 \choose k-1}
                     \sum^n_{a=1}\nu^2_a  \biggr]
        \ll \prod^{k+1} \Os \gg_0 \ll \Oi ~\prod^{n-k} \Os \gg_0
                = 0 ~,
           \nonumber
\eea
where
\bb
       \lambda_1 = \sq \pi \kappa^2 ~.
\ee
This equation is valid for $n \geq 3$. The $n=2$ equation is trivially
satisfied due to the vanishing of the corrlators with odd numbers
of $\Os$. The $n=1$ is ill-defined as discussed in the last of Sect.2.
The factorization form is given as follows. The sum over $k$ stands for
choosing $k$ operators which form the set $X$ out of $n$ $\Os$'s in the
set $S$. Then we have the following formulas:
\bb
     \sum_{S=X \cup Y} = \sum^n_{k=1} \sum_{\{ X \}_k} ~,
\ee
and
\bba
   &&  \sum_{\{ X \}_k} 1 = {n \choose k} ~, \qquad
       \sum_{\{ X \}_k} \sum_{a \in X} \nu_a = {n-1 \choose k-1} \chi ~,
                \nonumber         \\
   &&  \sum_{\{ X \}_k} \biggl( \sum_{a \in X} \nu_a \biggr)^2
           = {n-2 \choose k-2} \chi^2
               +{n-2 \choose k-1}\sum^n_{a=1} \nu^2_a ~,
\eea
where $\sum_{\{ X \}_k}$ is taken over all possibility of choosing
$k$ out of $n$ $\Os$'s.  Using eq.(4.35) and above
formulas for sphere $\chi = \sum^n_{a=1} \nu_a =2$, we obtain the
expression (4.36).

   Let us next consider the the case of ${\cal O}=\Os \prod^n \Oi$.
Here we assign the curvatures only on $n$ $\Oi$ operators. Then we
get
\bba
   && \pi^2 \kappa^2  \sum^n_{k=0} \biggl[
          2 {n-2 \choose k} + \fr{8}{9}{n-2 \choose k-2}
           -\fr{1}{3} {n-2 \choose k-1}
                         \\
   && \quad
        + \fr{25}{18} {n-2 \choose k-1} \sum^n_{a=1}\nu^2_a
          \biggr] \ll \Os\Os \prod^k \Oi \gg_0
              \ll \prod^{n-k+1} \Oi  \gg_0 = 0 ~,
                 \nonumber
\eea
where we use $\ll \Oep \prod^n \Oi \gg_0 =0$.
This equation is valid for $n \geq 3$. This equation is, however,
rather trivial. For $n<3$ the correlator
$\ll W_{\s}~{\cal O} \gg_0 $ becomes ill-defined.

\subsection{Ward identities for $W_{\ep}$}
\indent

  In this subsection we consider the Ward identities for the current
$W_{\ep}(z,\z)$. The OPE between $W_{\ep}(z,\z)$ and the scaling
operators which contributes to the Ward identities is
\bba
   &&  dz W_{\ep}(z,\z) \Oi (w,\w)
         \nonumber     \\
   && \qquad  = d \zp \biggl[
          R^{\pp}_{\ep}(\zp)
           +\fr{\nu}{2}\fr{1}{\zp -\wp}S^{\pp}_{\ep}(\zp)
           - \fr{5\nu^2 +17\nu}{36(\zp -\wp)^2} T^{\pp}_{\ep}(\zp)
                 \biggr]
               \nonumber \\
   && \qquad\qquad\qquad\qquad\qquad \times
        {\bar B}^{\pp}_{\ep}(\z^{\pp})\Oi^{\pp}(\wp, \w^{\pp})
             \nonumber   \\
   && \qquad  = d \zp \fr{1}{\zp-\wp}
           \biggl( - \fr{3}{16\sq}\biggr)
            \biggl( 1+\fr{11}{18}\nu-\fr{55}{18} \nu^2 \biggr)
              \Oep^{\pp}(\wp, \w^{\pp}) ~,
\eea
where $\nu$ is the curvature singularity assigned on $\Oi$.

  We also need to calculate the following OPE that $W_{\ep}(z,\z)$
and two $\Os$'s get together and produce $\Oi$:
\bb
     dz  W_{\ep}(z,\z) \Os(0,0) \int d^2 w V_{\s}(w,\w)
          = d\zp \fr{1}{\zp} C(\nu_a,\nu_b) \Oi^{\pp} (0,0)~,
\ee
where $\nu_a$ and $\nu_b$ are the curvature assigned on two $\Os$'s.
The coefficient $C(\nu_a,\nu_b)$ are calculated as follows.
For the holomorphic part we get
\bba
  &&   dz \Rep(z) \Os(0) ~dw V_{\s}(w)
              \nonumber \\
  &&   = d\zp \biggl[   \Rep^{\pp} (\zp)
            + \half \biggl( \fr{\nu_a}{\zp}
                  + \fr{\nu_b}{\zp-\wp} \biggr) S^{\pp}_{\ep}(\zp)
              \nonumber  \\
  &&  \qquad
             -\fr{1}{36} \biggl(
               \fr{5\nu^2_a + 17\nu_a}{z^{\pp 2}}
               + \fr{5\nu^2_b + 17\nu_b}{(\zp-\wp)^2}
               + \fr{10\nu_a \nu_b}{\zp(\zp-\wp)}
                   \biggr) T^{\pp}_{\ep}(\zp) \biggr]
               \nonumber  \\
  &&  \qquad\qquad\qquad \times
         \Os^{\pp}(0)~ d\wp V^{\pp}_{\s}(\wp) ~,
\eea
where, to go to the non-singular frame, we use transformation law of
$\Rep(z)$ (4.13) and the coordinate transformation defined by
\bb
      \fr{\pd z}{\pd \zp}
        = \bigl( \zp -\zp_a \bigr)^{-\nu_a}
            \bigl( \zp -\zp_b \bigr)^{-\nu_b}
\ee
with $ \zp_a = 0$ and $\zp_b = \wp$.
After contracting all operators we get
\bba
    &&  \fr{1}{24\sq} d\zp  \biggl[
             \bigl(30-22\nu_a -55\nu^2_a \bigr)
                    \fr{1}{z^{\pp 2}}
          +  \bigl(30-22\nu_b -55\nu^2_b \bigr)
                    \fr{1}{(\zp-\wp)^2}
                 \nonumber  \\
    && \qquad\qquad\qquad\quad
          + \bigl( -150 + 165 \bigl(\nu_a +\nu_b \bigr)
                    -110 \nu_a \nu_b \bigr)
              \fr{1}{\zp(\zp-\wp)} \biggr]
                  \nonumber  \\
    && \qquad\qquad\qquad \times
           z^{\pp \fr{1}{3}} (\zp-\wp)^{\fr{1}{3}}
                w^{\pp -\fr{2}{3}} ~ d\wp
              d_{\ep\s\s} \Oi^{\pp} (0) ~,
\eea
where $C_{\ep\s\s}= d_{\ep\s\s} {\bar d}_{\ep\s\s}$. The calculation of
the anti-holomorphic part is rather simple because ${\bar \Bep}(\z)$ is
the BRST invariant operator with weight $0$ such that it is transformed
as a scalar. The result is
\bba
    && {\bar \Bep}(\z) \Os (0) ~d{\bar w} V_{\s}(\w)
             \nonumber  \\
    &&   =  \fr{1}{{\bar \zp}-{\bar \wp}}
               {\bar z}^{\pp \fr{1}{3}}
              ( {\bar \zp}-{\bar \wp} )^{\fr{1}{3}}
               {\bar w}^{\pp -\fr{2}{3}} ~d{\bar \wp}
                 {\bar d}_{\ep\s\s} \Oi^{\pp} (0) ~.
\eea
Combining the holomorphic and the anti-holomorphic parts and
carrying out the $w$-integral, we get
\bba
    && C(\nu_a,\nu_b) =
           -\fr{5}{\sq} \pi \biggl(
             \fr{\Gamma(\fr{1}{3})}{\Gamma(\fr{2}{3})} \biggr)^3
              C_{\ep\s\s}  \biggl[ 1 - \fr{77}{60} (\nu_a +\nu_b )
             \nonumber \\
    && \qquad\qquad\qquad\qquad\qquad\qquad\qquad
           + \fr{11}{48} \bigl( \nu^2_a +\nu^2_b \bigr)
           + \fr{11}{12}\nu_a \nu_b  \biggr]
             ~.
\eea
To derive this expression we use the following integral formula:
\bba
     I_n  &=& \int d^2 y |y|^{-\fr{4}{3}}
                  |1-y|^{-\fr{4}{3}} (1-y)^n
               \nonumber  \\
          &=& \pi \fr{\Gamma(\fr{1}{3})^2 \Gamma(\fr{1}{3}+n)}
                    {\Gamma(\fr{2}{3})^2 \Gamma(\fr{2}{3}+n)} ~.
\eea

   The anomalous correlator including the operator
${\bar \pd}W_{\ep}(z,\z)$ is evaluated as in the same way discussed
in Sect.4.1. Here we do not repeat the calculation. The result is
\bba
  &&  \ll \int d^2 z {\bar \pd} W_{\ep}(z,\z)
                      \prod_{a \in S} O_a \gg_0
              \nonumber  \\
  && = \fr{\pi^2 \kappa^2}{2} \sum_{S=X \cup Y}
            C_{\ep\s\s} \fr{1}{48}  \Bigl[
               1 - 55 \bigl(\nu_p -1 \bigr)^2 \Bigr]
              \nonumber   \\
  && \qquad\qquad\qquad \times
            \ll \Os ~\prod_{b \in X} O_b \gg_0
             \ll \Os ~\prod_{c \in Y} O_c \gg_0 ~,
\eea
where $\nu_p$ is the induced curvature which satisfies eq.(4.35).

   Let us consider the Ward identity of the type:
${\cal O}= \prod^n \Os $. Then the curvature $\nu_a ~(a=1, \cdots n)$
are assigned on $n$ $\Os$'s $(n \geq 1)$. Using the results calculated
above we get the following equation:
\bba
   && \pi \fr{3}{16\sq} \mu \ll \Oep ~\prod^n \Os \gg_0
             \nonumber  \\
   && -\fr{\pi}{\sq}\lambda_2  C_{\ep\s\s}
         \biggl[ \half n(n-1) -\fr{77}{30}(n-1) + \fr{11}{6}
                 \\
   && \qquad\qquad\qquad
            +\fr{11}{48}(n-3) \sum^n_{a=1} \nu^2_a
                \biggr] \ll \Oi \prod^{n-2} \Os \gg_0
              \nonumber  \\
   && -\fr{\pi}{2\sq} \lambda_1 C_{\ep\s\s}
           \sum^n_{k=0} \biggl[ \fr{54}{48}{n \choose k}
               -\fr{55}{48} {n-2 \choose k-1}
                \Bigl( 4- \sum^n_{a=1} \nu^2_a \Bigr)
                  \biggr]
                       \nonumber  \\
   && \qquad\qquad\qquad\qquad \times
           \ll \prod^{k+1} \Os \gg_0
             \ll \prod^{n-k+1} \Os \gg_0 =0  ~,
             \nonumber
\eea
where $\lambda_1$ is defined by (4.37) and
\bb
       \lambda_2 = 5\pi \biggl(
           \fr{\Gamma(\fr{1}{3})}{\Gamma(\fr{2}{3})}
                 \biggr)^3 ~.
\ee
The first term comes from the OPE between the current and the
potential $\Oi$.
To derive the second expression we use the relations
$\sum^n_{a=1} \nu_a =2$ and  \break
$\sum^n_{a,b =1 ~(a \neq b)}\nu_a \nu_b = 4- \sum^n_{a=1} \nu^2_a$.
This equation is defined for $n \geq 1$, but it becomes trivial
for odd $n$ due to the vanishing of correlators including the odd
numbers of $\Os$ by the fusion rules.

   The above equation and eqs.(4.36) and (4.40) have already derived
in the previous paper~\cite{ha}. Here we make clear the derivation of
these equations and give the explicit calculation of (4.47).
If ${\cal O}$ includes $\Oep$ we have to add the following type of
boundary contribution:
\bba
    && \lim_{\tau \rightarrow \infty} \kappa^2 \sum_{\D}
         \int^{\infty}_{-\infty} \fr{dp}{2\pi}
            \ll {\cal O}_1
               \nonumber  \\
    && \quad \times
         \biggl\{  \int_{\e^{-\tau} \leq |z| \leq 1}
                      d^2 z {\bar \pd}W_{\ep}(z,\z)
        \int_{|w| \leq |z|} d^2 w V_{\ep}(w,\w)
                 \\
    && \qquad\qquad
           + \int_{\e^{-\tau} \leq |z| \leq 1}
                      d^2 z  V_{\ep}(z,\z)
                \int_{|w| \leq |z|} d^2 w {\bar \pd}W_{\ep}(w,\w)
                \biggr\}   \nonumber \\
    &&\quad \times
            \fr{-2\pi}{H} \e^{-\tau H} |-p,\D \gg_0
                 \ll p, \D | ~{\cal O}_2 \gg_0 ~.
               \nonumber
\eea
This contribution is calculated by using the technique developed in
(4.24) and (4.42).  As a result we must add the following term to
r.h.s. of eq.(4.49):
\bba
   && -\fr{5}{2} \pi^3 \kappa^2  \biggl(
        \fr{\Gamma(\fr{1}{3})}{\Gamma(\fr{2}{3})} \biggr)^3
          \biggl[ 1 - \fr{11}{6}\nu_p  - \fr{209}{150}\nu
                  + \fr{11}{12} \nu^2_p   \\
   && \qquad
                  + \fr{11}{30} \nu^2
                   + \fr{11}{12} \nu_p \nu \biggr]
                  \ll {\check {\cal O}}_1 \Oi \gg_0
                  \ll \Oi ~{\check {\cal O}}_2 \gg_0 ~.
              \nonumber
\eea
Here $\nu$ is the curvature assigned on the operator $\Oep$ and
$\nu_p$ is that induced at the state $|-p,\D \gg$.
The check symbol $\vee$ on ${\cal O}_1$ and ${\cal O}_2$ stands
for the exclusion of the operator $\Oep$.
The $\Gamma$-functions come from the $w$-integrals. The integrals
of $p$ and $z$ are evaluated by using the saddle point method.

  Let us consider the case of ${\cal O}=\Oep \prod^n \Oi $.
The curvatures are assigned only on $n ~\Oi$'s $(n \geq 1)$.
Then, using the result (4.53) with $\nu =0$, we get~\footnote{
If we assign the curvature only on $\Oep$, we then get
\bba
    && \pi \fr{3}{16\sq} \mu \ll \Oep\Oep \prod^n \Oi \gg_0
        -\pi \fr{3}{16\sq} n \ll \Oep\Oep \prod^{n-1} \Oi \gg_0
             \nonumber  \\
    && \qquad
         + \pi\fr{2}{25\sq}\lambda_3 \sum^n_{k=0}
              {n \choose k} \ll \prod^{k+1} \Oi \gg_0
                   \ll \prod^{n-k+1} \Oi \gg_0   = 0
\eea
for $n \geq 0$, where we use the formula (4.53) with $\nu=2$ and $\nu_p = 0$.
} 
\bba
    && \pi \fr{3}{16\sq} \mu \ll \Oep \Oep \prod^n \Oi \gg_0
              \nonumber  \\
    &&   -\pi \fr{3}{16\sq} \biggl(
                  n+\fr{11}{9} -\fr{55}{18} \sum^n_{a=1} \nu^2_a
           \biggr) \ll \Oep \Oep \prod^{n-1} \Oi  \gg_0
                   \nonumber  \\
    &&  -\fr{\pi}{4\sq} \lambda_3
           \sum^n_{k=0} \biggl[ {n \choose k}
              - \fr{11}{12} {n-2 \choose k-1}
                 \Bigl( 4- \sum^n_{a=1} \nu^2_a  \Bigr) \biggr]
                       \\
    && \qquad\qquad \times
              \ll \prod^{k+1} \Oi \gg_0
                  \ll \prod^{n-k+1} \Oi  \gg_0 = 0 ~,
                    \nonumber
\eea
where
\bb
      \lambda_3 = \half 4\sq \fr{5}{2} \pi^2 \kappa^2
                    \biggl(
           \fr{\Gamma(\fr{1}{3})}{\Gamma(\fr{2}{3})}
                     \biggr)^3
               = \lambda_1 \lambda_2  ~.
\ee
The factor $\half$ in the middle of eq.(4.56) corrects for double
counting.

  If we take ${\cal O}= \Oep\prod^n \Os$ and assign the curvatures on
$n$ $\Os$'s, we get
\bba
     && \pi \fr{3}{16\sq} \mu \ll \Oep\Oep \prod^n \Os \gg_0
               \nonumber  \\
     && -\fr{\pi}{\sq} \lambda_2 C_{\ep\s\s} \sum^n_{k=0}
            \biggl[ \half n(n-1) -\fr{77}{30}(n-1)
                 +\fr{11}{6} +\fr{11}{48}(n-3) \sum^n_{a=1} \nu^2_a
                      \biggr]
               \nonumber  \\
     && \qquad\qquad\qquad\qquad\qquad\qquad\qquad \times
              \ll \Oi\Oep \prod^{n-2} \Os \gg_0
                \nonumber \\
     && -\fr{\pi}{\sq} \lambda_1 C_{\ep\s\s} \sum^n_{k=0}
            \biggl[ \fr{54}{48} {n \choose k}
                - \fr{55}{48} {n-2 \choose k-1}
                       \Bigl( 4-\sum^n_{a=1} \nu^2_a \Bigr) \biggr]
                            \\
     && \qquad\qquad\qquad\qquad\qquad \times
           \ll \Oep \prod^{k+1} \Os \gg_0
                    \ll \prod^{n-k+1} \Os \gg_0
                  \nonumber \\
     && - \fr{\pi}{4\sq} \lambda_3 \sum^n_{k=0}
             \biggl[ {n \choose k} -\fr{11}{12}{n-2 \choose k-1}
                       \Bigl( 4-\sum^n_{a=1} \nu^2_a \Bigr) \biggr]
                  \nonumber  \\
     && \qquad\qquad\qquad\qquad\qquad \times
              \ll \Oi \prod^k \Os \gg_0
                 \ll \Oi \prod^{n-k} \Os \gg_0 = 0 ~.
               \nonumber
\eea
This equation is valid for $n \geq 4$. $n=2$ is ill-defined and
$n= \mbox{odd}$ is trivial.

  Until now we introduce the three independent parameters, $\mu$ and
two of $\lambda_j ~(j=1,2,3)$, which are related to the normalizations of
the three scaling operators. The normalization of $\Oi$ is fixed by the
equation $ -\fr{\pd}{\pd \mu} \ll {\cal O} \gg_0
= \ll \Oi~{\cal O} \gg_0 $.
If we change the normaizations of $\Os$ and $\Oep$ into
${\tilde \Os}= x \Os$ and ${\tilde \Oep}=y \Oep$, and
simultaneously $\lambda_j$ into
${\tilde \lambda}_1 =(y/x^2) \lambda_1$,
${\tilde \lambda}_2 = x^2 y \lambda_2$ and
${\tilde \lambda}_3 = y^2 \lambda$, the recurion relations
do not change the forms and they also satisfy the relation
${\tilde \lambda}_3 = {\tilde \lambda}_1 {\tilde \lambda}_2 $.

\section{Solutions of Recursion Relations}
\setcounter{equation}{0}
\indent

    In this section we give the several solutions of the recursion
relations (4.36), (4.50), (4.55) and (4.57) and compare them with the
results of the matrix model.
These equations are satisfied independently of the
value of $\sum^n_{a=1} \nu^2_a $ for $n \geq 2$ because how to assign
the curvature singularities is completely arbitrary.  For $n=1$, however,
it is fixed to the value $\sum_a \nu^2_a = \nu^2_1 =4$.

   For simplicity we here use the following notation:
\bb
     \ll \prod^{n_1} \Oi \prod^{n_2} \Os \prod^{n_3} \Oep \gg_0
       = < \prod^{n_1} I \prod^{n_2} \s \prod^{n_3} \ep > ~.
\ee
Let us write down the first several of eq.(4.36).
{}From $n=3$  we get the following two equations:
\bba
      5 C_{\ep\s\s} < \ep\s\s >
        -3 \lambda_1 < \s\s > < \s\s I >
          + 4 \lambda_1 < \s\s\s\s >< I >
         & = & 0
                  \\
      C_{\ep\s\s} < \ep\s\s >
        - \lambda_1 < \s\s >< \s\s I > &=& 0 ~.
\eea
The second comes from the coefficients in front of
$\sum_a \nu^2_a$. The $n=4 $ equation is now trivial. The $n=5$ equation
gives
\bba
    && 15 C_{\ep\s\s} <\ep\s\s\s\s> + 15\lambda_1 <\s\s><\s\s\s\s I>
                      \\
    && \qquad
        -15\lambda_1 <\s\s\s\s><\s\s I> + 4\lambda_1 <\s\s\s\s\s\s><I>
          = 0
               \nonumber
\eea
and, from the coefficient in front of $\sum_a \nu^2_a$,
\bb
    -C_{\ep\s\s} <\ep\s\s\s\s> +\lambda_1 <\s\s><\s\s\s\s I>
          + 3\lambda_1 <\s\s\s\s><\s\s I> = 0 ~.
\ee

  Next we write down the first several of eq.(4.50). From $n=2$,
we obtain the following two equations:
\bba
      \fr{3}{16} \mu <\ep\s\s>
       - \fr{4}{15} \lambda_2 C_{\ep\s\s} <I>
       + \fr{7}{6} \lambda_1 C_{\ep\s\s} <\s\s>^2  &=& 0 ~,
                   \\
    2 \lambda_2 <I> - 5 \lambda_1 <\s\s>^2 &=& 0 ~.
\eea
{}From $n=4$ we get
\bb
     \fr{3}{16}\mu <\ep\s\s\s\s>
      -\fr{2}{15} \lambda_2 C_{\ep\s\s} <\s\s I>
      + \fr{1}{12} \lambda_1 C_{\ep\s\s} <\s\s><\s\s\s\s>
         = 0
\ee
and
\bb
       \lambda_2 <\s\s I> + 5 \lambda_1 <\s\s><\s\s\s\s> = 0 ~.
\ee

  Furthermore, from $n=1$ of (4.55), we obtain~\footnote{
This equation is the same as the $n=0$ equation of (4.54) essentially.
}
\bb
     -3 \mu <\ep\ep I>  -30 <\ep\ep>
          + 8 \lambda_3 <I><II> = 0   ~.
\ee

  Noting that $< I ~{\cal O} > = -\fr{\pd}{\pd \mu}
< {\cal O} >$ such that
$ < \s\s > = -(3/2)\mu < \s\s I >~, \quad < 1 > = -(3/7) \mu < I >
= (9/28) \mu^2 < I I > ~,\quad <\ep\ep>=-\fr{3}{5}\mu<\ep\ep I> $
and so on, we obtain the folllowing normalization independent ratio:
\bb
     \fr{< \ep\s\s >^2 < 1 >}
          {< \ep\ep >< \s\s >^2}
         = \fr{20}{7} \fr{\lambda_1 \lambda_2}{\lambda_3}
                         C^2_{\ep\s\s}
         = \fr{5}{7} ~.
\ee
Here we use eqs.(5.6), (5.7), (5.10) and the relation (4.56) and
$C^2_{\ep\s\s}=\fr{1}{4}$.  The same result is also derived by using
eqs.(5.3) instead of (5.6). It is just a consistency check of the result.
This value agrees with that derived by the other
methods~\cite{dfk,cgm,gl}. From (5.9) and (5.7) we get
\bb
     \fr{< \s\s\s\s >< 1 >}
          {< \s\s >^2}
       = -\fr{1}{7} ~.
\ee
It is easily checked that the same result is given by using (5.2), (5.3)
and (5.7). This also agrees with the result derived by
Crnkovi\'{c} et. al.~\cite{cgm}.
Furthermore we get following values:
\bb
      \fr{<\ep\s\s\s\s>^2 <1>^3 }{<\ep\ep><\s\s>^4 }
          = \fr{45}{1372}
\ee
by using eq.(5.5) or (5.8) and
\bb
      \fr{<\s\s\s\s\s\s><1>^2 }{<\s\s>^3 }
          = -\fr{15}{98}
\ee
by using eq.(5.4).

  {}From $n=4$ of eq.(4.57) we get
\bba
   &&  \fr{3}{16}\mu <\ep\ep\s\s\s\s>
       -\fr{2}{15}\lambda_2 C_{\ep\s\s}<\ep\s\s I>
            \nonumber \\
   &&   +\fr{1}{12}\lambda_1 C_{\ep\s\s} <\ep\s\s><\s\s\s\s>
        + \fr{1}{12}\lambda_1 C_{\ep\s\s} <\ep\s\s\s\s><\s\s>
              \nonumber  \\
   &&   -\half \lambda_3 <\s\s\s\s I><I>
        +\fr{1}{3} \lambda_3 <\s\s I>^2  = 0 ~.
\eea
The equation coming from the coefficients in front of $\sum_a \nu^2_a$
does not give any new information, which is automatically satisfied
by using the equations listed above. From (5.15) we obtain
\bb
     \fr{<\ep\ep\s\s\s\s><1>^2 }{<\ep\ep><\s\s>^2 }
          = - \fr{15}{98} ~.
\ee

\section{Summary and Discussion}
\setcounter{equation}{0}
\indent

    In this paper we discussed the recursion relations of the Liouville
gravity coupled to the non-bosonized Ising matter satisfying the
fusion rules and found out a closed set of them. They
have three independent parameters, $\mu$ and two of
$\lambda_j ~(j=1,2,3)$ satisfying the relation
$\lambda_3 = \lambda_1 \lambda_2$, which are related to the normalizations
of the three scaling operators $\Oi$, $\Os$ and $\Oep$. We obtained some
solutions and computed the normalization independent ratios. The results
agree with those calculated in the other methods~\cite{cgm,dfk,gl}.

   The difference from the theory using the bosonized matter is whether
the gravitational descendants other than $\s_1 (\Os)=\Oep$ appear in the
recursion relations or not.
If we replace $\s$ and $\ep$ in (2.4), (4.1-2) and (4.4-5) into the
bosonized forms\footnote{
In the bosonized theory there exist the physical operators satisfying
the $W_{\infty}$ algebra as for $R(z)$~\cite{w}. So there is no reason to
use the operators of type listed in (4.4-5) in the bosonized theory.
} 
and introduce a screening charge in the definition of the
correlation functions as in ref.~\cite{hb}, $\s_1 (\Oi)$  will appear and
some of the OPE coefficients are changed~\footnote{
If we use the $W_{\infty}$ currents with the same Liouville charges as
(4.4-5) instead, we furthermore encounter the gravitational descendants
$\s_3 (\Oi)$ and $\s_3 (\Os)$ because the OPE's between the currents and
the screening charge operator do not vanish~\cite{hb}.
}. 
 Conversely the disappearance of these operators indicates that the
fusion rules are satisfied.

   In exchange for the disappearance of the gravitational descendants
other than $\Oep$, we encounter the demerit that the correlators
\bba
  &&   \ll \Oep \gg_0 ~, \quad \ll \Oep \Oi \gg_0 ~, \quad
       \ll \Oep\Oi\Oi \gg_0 ~,
               \\
  &&   \ll \Oep\Oep\Oep\Oep \gg_0 ~, \quad
       \ll \Oep\Oep\Oep\Oep\Oi \gg_0 ~, \quad
       \ll \Oep\Oep\Os\Os \gg_0 ~,
            \nonumber
\eea
which have zero or positive integer $s$, and
also $\ll \Oep\Oep\Oep \gg_0$, whose Liouville part diverges, become
ill-defined. When the recursion relations include such a correlator they
become inconsistent, or there is no solution satisfying them independently
of how to assign the curvature singularities.
This make impossible to calculate the
correlators including more than and equal three $\Oep$'s.

  We do not know how to regularize these correlators. The attempt to
determine the values of these correlators by imposing the requirement
that the equations are naively extended to the case which include them,
for instance $n=1$ of eq.(4.36), does not do well. There is no solutions
satisfying such a requirement.

   To make these correlators well-defined we have to go to the
bosonized theory perturbed by the cosmological constant operator and
{\it one} of the screening charges of matter sector~\cite{hb}. Then
the factor $\Gamma(-s)/n! =\Gamma(-s)/\Gamma(n+1)$ satisfying the relation
$s+n+N=\chi=2$ appears in the definition of correlator, where
$\Gamma(-s)$ and $1/n!=1/\Gamma(n+1)$
come from the zero mode integrals of the Liouville
and the matter fields respectively and $N$ is the number of the scaling
operators in $\ll \cdots \gg_0$. So the divergence of $\Gamma(-s)$ at
$s \in {\bf Z}_{\geq 0}$ cancels out with $\Gamma(n+1)$ in the case listed
in (6.1).
The correlator $\ll \Oep\Oep\Oep \gg_0 $ become finite in rather different
way as discussed by Kitazawa in ref.~\cite{gl}.

\end{document}